\documentclass[%
 reprint,
 amsmath,amssymb,superscriptaddress,
 aps,
]{revtex4-2}

\usepackage{physics,color}
\usepackage{graphicx}
\usepackage{dcolumn}
\usepackage{bm}
\usepackage{ulem}

\begin{document}

\preprint{APS/123-QED}

\title{A scheme for multipartite entanglement distribution via separable carriers}

\author{Alessandro Laneve}
\email{alessandro.laneve@uniroma1.it}
\affiliation{Dipartimento di Fisica, Sapienza Universit\`{a} di Roma, Piazzale Aldo Moro, 5, I-00185 Roma, Italy}
\author{Hannah McAleese}
\affiliation{Centre for Theoretical Atomic, Molecular, and Optical Physics, School of Mathematics and Physics, Queen’s University Belfast, BT7 1NN Belfast, UK}
\author{Mauro Paternostro}
\affiliation{Centre for Theoretical Atomic, Molecular, and Optical Physics, School of Mathematics and Physics, Queen’s University Belfast, BT7 1NN Belfast, UK}

\date{\today}

\begin{abstract}
The ability to reliably distribute entanglement among the nodes of a network is an essential requirement for the development of effective quantum communication protocols and the realization of useful quantum networks. 
It has been demonstrated, in different contexts, that two remote systems can be entangled via local interactions with a carrier system that always remains in a separable state with respect to such distant particles.  We develop a strategy for entanglement distribution via separable carriers that can be applied to any number of network nodes to achieve various entanglement distribution patterns. We show that our protocol results in multipartite entanglement, while the carrier mediating the process is always in a separable state with respect to the network. We provide examples showcasing the flexibility of our approach and propose a scheme of principle for the {experimental} demonstration of the protocol.
\end{abstract}

\maketitle


\section{Introduction}

The crucial role played by entanglement in quantum networking and communication has long been established through a plethora of groundbreaking protocols and experimental demonstrations \cite{ekert1992quantum,bennett1993teleporting,cirac1997quantum, boschi1998experimental,bouwmeester1997experimental, ursin2007entanglement, humphreys2018deterministic,wengerowsky2018entanglement}. It is therefore imperative that efficient methods of entanglement generation among the nodes of a quantum network are developed. In particular, we need protocols which take the fragility of entanglement into account, for instance, by creating this vital resource just before it is needed to be used. 

Say we have two parties, Alice and Bob, who aim to share entanglement. To distribute entanglement directly, Alice would create an entangled state of two particles in her laboratory before sending one particle to Bob through a quantum channel. Alternatively, Alice and Bob could distribute entanglement indirectly through the use of an ancilla. This carrier system would first interact with Alice's particle in her laboratory, then be sent to Bob. This process will often require the carrier to become entangled with the two systems.

However, it is possible to entangle Alice and Bob's systems in this way without ever entangling either system with the ancilla. Theoretical proposals were put forward for entanglement distribution via separable states (EDSS) in the discrete-variable case~\cite{cubitt2003separable,kay12using} and continuous-variable case~\cite{mista08distribution,mista09improving,mista13entanglement} before it was demonstrated experimentally~\cite{fedrizzi2013experimental,vollmer13experimental,peuntinger13distributing}. It is important to note that quantum discord is necessary for EDSS to be possible~\cite{chuan2012quantum}. As discord is much more robust to noise and environmental effects than entanglement~\cite{werlang09robustness,ferraro10almost,wang10nonmarkovian,fanchini10nonmarkovian,mazzola10sudden}, EDSS provides an advantage over protocols which rely on the presence of entanglement. 
Interestingly, the fact that EDSS depends on non-classical correlations also means it can be used to detect non-classicality in inaccessible objects~\cite{krisnanda17revealing,krisnanda18probing,krisnanda19observable}.

In what follows, we will build on Kay's EDSS protocol for qubits~\cite{kay12using}. The procedure is as follows: Firstly, Alice and Bob initially share a separable state of their systems $A$ and $B$. Secondly, Alice introduces an ancilla system $K$ which is uncorrelated from $AB$. Thirdly, Alice performs the {\it encoding operation}, that is, a unitary operation $U_{AK}$ on her system and the carrier. Finally, Alice sends $K$ to Bob. Ref.~\cite{kay12using} shows that when $AB$ is initially in a Bell-diagonal state and $U_{AK}$ is a controlled-phase gate, it is possible to choose a suitable initial state for $K$ so that the state of the total system at the end of the protocol is entangled in the bipartition $A|BK$ and $K$ remains separable from $A$ and $B$ throughout the process. In what follows, we add an extra step to the protocol; Bob performs a {\it decoding operation} on his particle and the carrier after he receives $K$ from Alice. This results in entanglement in both the $A|BK$ and $B|AK$ bipartitions while the ancilla remains separable with no entanglement in the partition $K|AB$. 

In this work, we generalize the protocol in Ref.~\cite{kay12using} to the distribution of multipartite entanglement through EDSS, specifically focusing on the conditions of its experimental demonstration in Ref.~\cite{fedrizzi2013experimental}. Multipartite EDSS has previously been addressed in Ref.~\cite{karimipour2015systematics}, where a systematic method was proposed based on the EDSS protocol by Cubitt et al.~\cite{cubitt2003separable}. In this case, $AB$ and $K$ are initially correlated (yet unentangled). The risk of entangling the bipartition $K|AB$ is therefore higher in their proposal and extra effort must be made to ensure its prevention. As the initial state of $K$ in Refs.~\cite{kay12using,fedrizzi2013experimental} shares no classical or non-classical correlations with $A$ or $B$, we avoid this problem and show that favouring this type of protocol offers a promising avenue for successful EDSS with fewer restrictions. 

The remainder of this paper is structured as follows: In Sec.~\ref{sec:protocol}, we introduce the protocol and show that the strategy relies on a particular initial state setting, that allows us to infuse the system with the initial quantum correlations necessary to obtain, at the end of the protocol,  multipartite entanglement. In Sec.~\ref{sec:analysis} we show that this protocol can be applied to different entanglement distribution patterns and that, in general, it represents a very flexible approach to the problem of 
EDSS. In addition, in Sec.~\ref{sec:experiment}, we propose two possible experimental platforms for the implementation of such a protocol in a photonic scenario. Finally in Sec.~\ref{sec:conclusions} we present our conclusions.

\section{Illustration of the protocol} \label{sec:protocol}
\subsection{Two-qubit Protocol}
\begin{figure}
    \includegraphics[width=0.85\columnwidth]{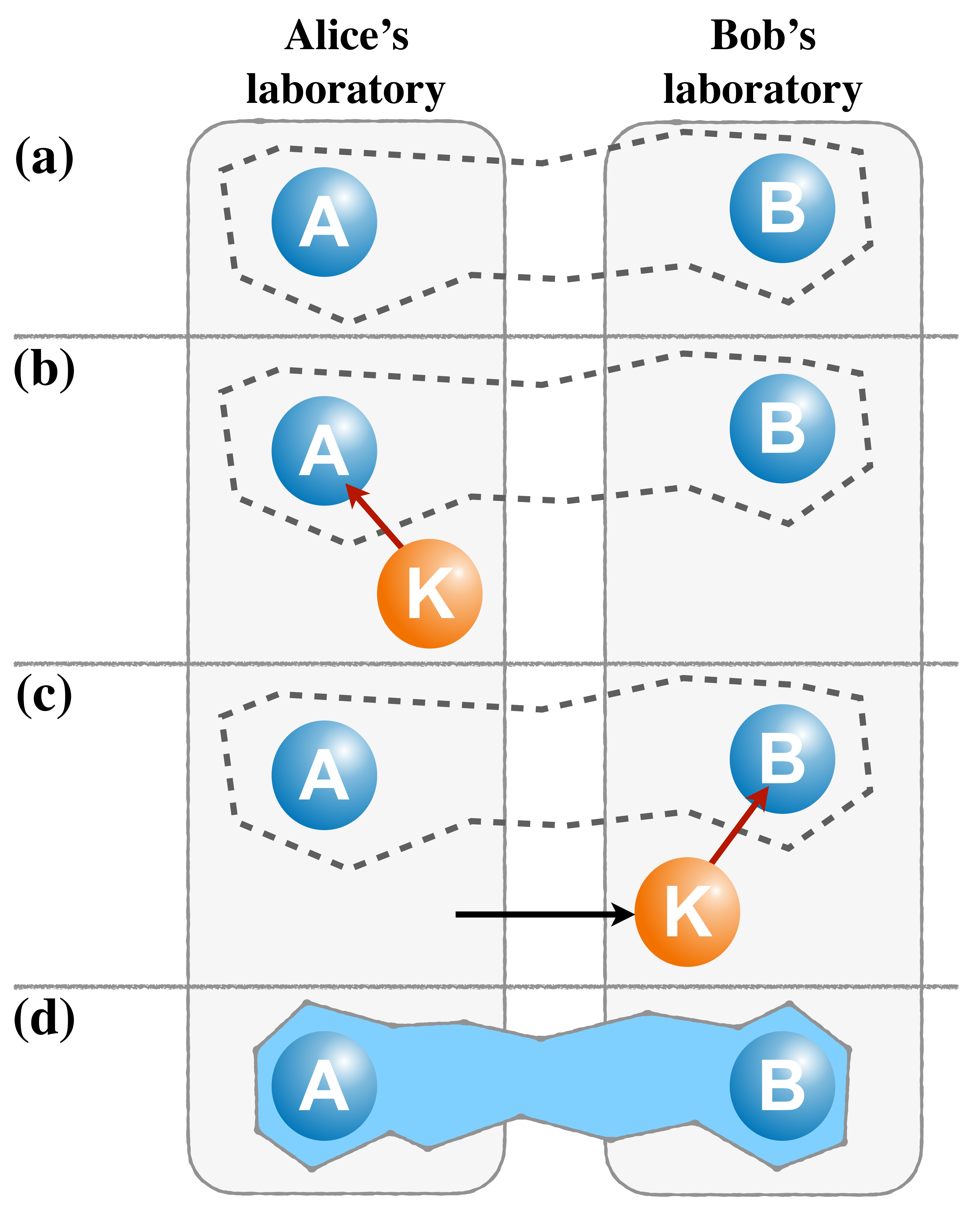}
    \caption{{\bf Diagram of the two-qubit protocol.} {\bf (a)} Nodes $A$ and $B$ are initially separable and they share quantum discord (dashed line). {\bf (b)} Alice introduces the carrier $K$ which is completely uncorrelated from $AB$. The encoding operation, in this case a controlled-phase gate, is performed between $A$ and $K$. {\bf (c)} Carrier $K$ is sent to Bob and $B$ and $K$ interact via the decoding operation. {\bf (d)} Particles $A$ and $B$ now share quantum entanglement.}
    \label{fig:protocol}
\end{figure}

We will refer explicitly to the version of the protocol for entanglement distribution with separable states that has been reported in  Ref.~\cite{fedrizzi2013experimental} as illustrated in Fig.~\ref{fig:protocol}. In such a scheme, the initial state of the two nodes $A$ and $B$ is separable, yet features non-classical correlations (as quantified by quantum discord~\cite{chuan2012quantum}). Explicitly, we take
\begin{equation}
\label{ini2}
    \begin{aligned}
     \alpha_{AB}&=\frac{1}{4}\left(\ket{00}\bra{00}+\ket{11}\bra{11}\right)+\frac{1}{8}\left(\ket{DD}\bra{DD}\right.\\
     &+\left.\ket{AA}\bra{AA}+\ket{RL}\bra{RL}+\ket{LR}\bra{LR}\right)_{AB},
\end{aligned}
\end{equation}
{
where $\ket{D}=\frac{1}{\sqrt{2}}(\ket{0}+\ket{1})$, $\ket{A}=\frac{1}{\sqrt{2}}(\ket{0}-\ket{1})$,  $\ket{R}=\frac{1}{\sqrt{2}}(\ket{0}+i\ket{1})$ and $\ket{L}=\frac{1}{\sqrt{2}}(\ket{0}-i\ket{1})$.}
While $\alpha_{AB}$ is invariant under partial transposition, it is endowed with non-zero quantum discord, as quantified by the relative entropy of discord~\cite{Modi2010,Synak2005}
\begin{equation}
{\cal D}(\alpha_{AB})=\min_{\Pi_B}[S(\Pi_B(\alpha_{AB}))]-S(\alpha_{AB}),
\end{equation}
where $S(\rho)$ is the von Neumann entropy of state $\rho$ and $\Pi_B(\rho)=\sum^1_{j=0}\pi_j\rho\pi_j$ is a rank-one projective measurement of $\rho$ with $\pi_{0,1}$ two orthogonal projectors on qubit $B$. We have ${\cal D}(\alpha_{AB})=0.0612781$.

The state of the nodes is then subjected to  encoding and decoding operations, each consisting of a controlled-phase (CPHASE) gate acting on the joint state of either $A$ or $B$ and the carrier $K$. The initial state of the latter must be a mixture of orthogonal vectors that are maximally distant from the eigenstates of $\sigma_z$, in order to amplify the effect of the encoding and decoding operations. We thus choose
\begin{equation}
\label{iniK}
\alpha_K=\frac{1}{4}\left(\ket{D}\bra{D}+{3}\ket{A}\bra{A}\right)_K,
\end{equation}
although a mixture of $\ket{R}$ and $\ket{L}$ would also be suitable.
These mixing probabilities are chosen so as to guarantee that the carrier is not entangled throughout the process, while achieving the largest possible entanglement between the two nodes at the end of the protocol. In this sense, a mixture with balanced probabilities would be suitable too, although less effective. 

The protocol now involves the encoding step, when the CPHASE gate is applied to qubit $A$ and the carrier. This is then followed by a decoding step, consisting of the application of the CPHASE to node $B$ and carrier. The first step reads
\begin{equation}
    \beta_{ABK}=\mathcal{P}_{AK}(\alpha_{AB}\otimes\alpha_K)\mathcal{P}_{AK}^\dag
\end{equation}
where
\begin{equation}
    \mathcal{P}_{AK}=\ket{0}\bra{0}_A\otimes \openone_K+\ket{1}\bra{1}_A\otimes \sigma_{z,K}
    \label{eq:gate_binary}
\end{equation}
is the CPHASE gate between $A$ and the carrier $K$.
The decoding step then gives
\begin{equation}
    \gamma_{ABK}=\mathcal{P}_{BK}\beta_{ABK}\mathcal{P}_{BK}^\dag.
\end{equation}
The resulting state $\gamma_{ABK}$ features distillable entanglement in the bipartitions $A|BK$ and $B|AK$ with the carrier $K$ being in a separable state with respect to the state of the nodes (either collectively or individually taken).

\subsection{General protocol}

We now exploit the same encoding and decoding mechanisms illustrated above to design a generalization of the two-qubit protocol to a multipartite set of nodes. The resource being exploited is a mixed state that features initial non-classical correlations between each of the node pairs. The scope of the process is to entangle the elements of the network.

\subsubsection{Initial state of the network, carrier state, and encoding-decoding operations}
We consider the case of a network of $N$ nodes $\{Q_i\}$~($i=1,\dots,N$) and investigate the arrangement of a protocol capable of establishing a pattern of entangled links between such nodes, according to a given structure. Thus, we require the definition of a state which features non-classical correlations between the nodes we wish to get entangled.
In order to do that, we use the two-qubit state featuring quantum discord that was employed in the two-qubit case of Eq.~\eqref{ini2}.

Proceeding in analogy with the bipartite case, we generalize this state to $N$ qubits by imposing a mixed initial state, consisting of a balanced mixture of terms featuring non-classicality between every  pair of nodes targeted by our protocol.
Each of such terms features a correlated state of a given  pair of nodes, while the other nodes are set in an eigenstate of the encoding and decoding operation.
We define a list of two-element sets containing the $M$  pairs we wish to entangle, labelling them as $\{\mathcal{C}_k\}_{k=1}^M$, where each $\mathcal{C}_k=\{Q_i , Q_j\}$ represents a different node pair $\{i,j\}$ among the chosen ones. The initial state of the network $\alpha_N$ has thus the form
\begin{equation}
 \alpha_N=\frac{1}{M}\sum_{k=1}^M \rho^0_{\mathcal{C}_k}\left(\bigotimes_{Q_i\notin {\cal C}_k}\alpha^0_{Q_i}\right),
\label{eq:initialAlphaN}
\end{equation}
where $\alpha^0_{Q_i}=\ket{0}\bra{0}_{Q_i}$ and $\rho^0_{\mathcal{C}_k}$ is the initial state in Eq.~\eqref{ini2}, but for the pair ${\cal C}_k$. 
For instance, we can choose to distribute entanglement according to a chain-like structure, namely a linear network in which each node is entangled with its closest neighbours. The initial state can be written in the compact form as
\begin{equation}
\label{newnetworkstate}
  \alpha_N^{linear}=\frac{1}{N-1}\sum_{k=1}^{N-1} \rho^0_{Q_k,Q_{k+1}}\left(\bigotimes_{Q_i\neq \{Q_k,Q_{k+1}\}}\alpha^0_{Q_i}\right).
\end{equation}
Such an initial state is necessary to keep the carrier in a separable state with respect to the network. Basically, in this way we are able to carry the two-qubit protocol in parallel over any node pair we want to entangle, without interference between the various terms.
As we will see later on, this has some interesting implications for the features of the final state.

The encoding and decoding operations consist of CPHASE gates ${\cal P}_{Q_iK}$ acting on the state of node $Q_i$ and the carrier $K$, 
whose initial state is chosen again as in Eq.~\eqref{iniK}.
Other equivalent gate-carrier initial state pairings exist, though they do not result in better performance of the protocol. 



\subsubsection{Single qubit carrier}

In this case, we only have one carrier $K$ and the total initial state of the system can be set as
\begin{equation}
    \alpha_T=\alpha_N\otimes\alpha_K
\end{equation}
so that the preparation of the network system and the carrier can be independently addressed.
As the carrier is the same for each pair of nodes, a single encoding and decoding step for each qubit is enough for weaving multiple entanglement links.
 The effect of the local CPHASE gate on the total state is 
\begin{equation}
  \mathcal{P}_{Q_l K}  \alpha_T\mathcal{P}_{Q_l K}^\dag=
    \frac{1}{M}\sum_{k=1}^M\chi_{k,Q_l}\left(\bigotimes_{Q_i\notin \mathcal{C}_k}\alpha^0_{Q_i}\right),
\end{equation}
where
\begin{equation}
    \chi_{k,Q_l}=
    \begin{cases}
    \rho^0_{\mathcal{C}_k}\otimes\alpha_K~~~~~~~~~~~~~~~~~~~~~\text{for $Q_l\notin \mathcal{C}_k$},\\
    \mathcal{P}_{Q_l K} \big(\rho^0_{\mathcal{C}_k}\otimes\alpha_K\big)\mathcal{P}_{Q_l K}^\dag~~~~\text{for $Q_l\in \mathcal{C}_k$}.
    \end{cases}
\end{equation}
The CPHASE gate on qubit $Q_l$ acts as an encoding operation on the terms involving $Q_l$ as a target or a decoding one, while acting as the identity on the others.

\subsubsection{Multiple qubit carriers}

It is possible to tailor the above protocol to work with multiple carriers. 
We also investigate this case in order to understand which beneficial effects and costs derive from this choice.
We consider 
compound of $n$ qubits $\{K_i\}_{i=1}^n$ and take as initial state for each the state $\alpha_K$ in Eq.~\eqref{iniK}. This sets the total product state of the carrier compound as
\begin{equation}
    \alpha_{\bar{K}}=\bigotimes_{i=1}^n\alpha_K^i=\bigotimes_{i=1}^n\left(\frac{1}{4}\ket{D}\bra{D}_{K_i}+\frac{3}{4}\ket{A}\bra{A}_{K_i}\right), 
\end{equation}
so that the total initial state is simply
\begin{equation}
    \alpha_T=\alpha_N\otimes\alpha_{\bar{K}}.
\end{equation}
The main difference with respect to the one qubit carrier protocol is that, in the present case, we proceed to entangle each qubit pair making them interact with a different carrier. Therefore, node qubit $Q_i$ is subject to encoding via the  carrier qubit $K_i$, which also mediates the local decoding at $Q_{i+1}$. Then, the encoding between $Q_{i+1}$ and $Q_{i+2}$ is mediated by  carrier $K_{i+1}$. Therefore,  different encoding and decoding operations are needed. 


In general, the final state of the nodes will have the form of a mixture of terms stemming from the various two-qubit processes, which could take place in parallel, and an incoherent term $\Omega_N$, thus reading
\begin{equation}
    \rho^f_N=P \Omega_N+(1-P)\sum_{k=1}^M \ket{\phi^+}\bra{\phi^+}_{\mathcal{C}_k}\left(\bigotimes_{Q_i\notin \mathcal{C}_k}\alpha^0_{Q_i}\right),
\end{equation}
where $\ket{\phi^+}=\frac{1}{\sqrt 2}(\ket{00}+\ket{11})$ is a Bell state and, as stated previously, $\{\mathcal{C}_k\}_{k=1}^M$ is the list of node pairs we aim to entangle. 
\textcolor{black}{The mixing coefficient $P$ is determined by the terms we insert in the initial mixed states, hence the number of entangled links we wish to establish. For each contribution, an incoherent residual term appears in the final state, forming the global incoherent term $\Omega_N$.}
Therefore, the final state is a mixture of terms featuring bipartite entanglement, one for each of the initially non-classically correlated node pairs. Clearly, that implies a probabilistic generation of entanglement.
Nevertheless, as we explicitly show in the following examples, the final state of the system unambiguously exhibit multipartite entanglement, namely the network is entangled with respect to any possible bipartition.

\section{Analysis of performance} \label{sec:analysis}

In this Section, we analyze the  performance of both single- and multiple-carrier protocols by addressing two explicit examples.

\subsection{Four nodes example: Ring configuration}

\subsubsection{Single carrier}

\begin{figure}
{\bf (a)}
    \includegraphics[width=0.65\columnwidth]{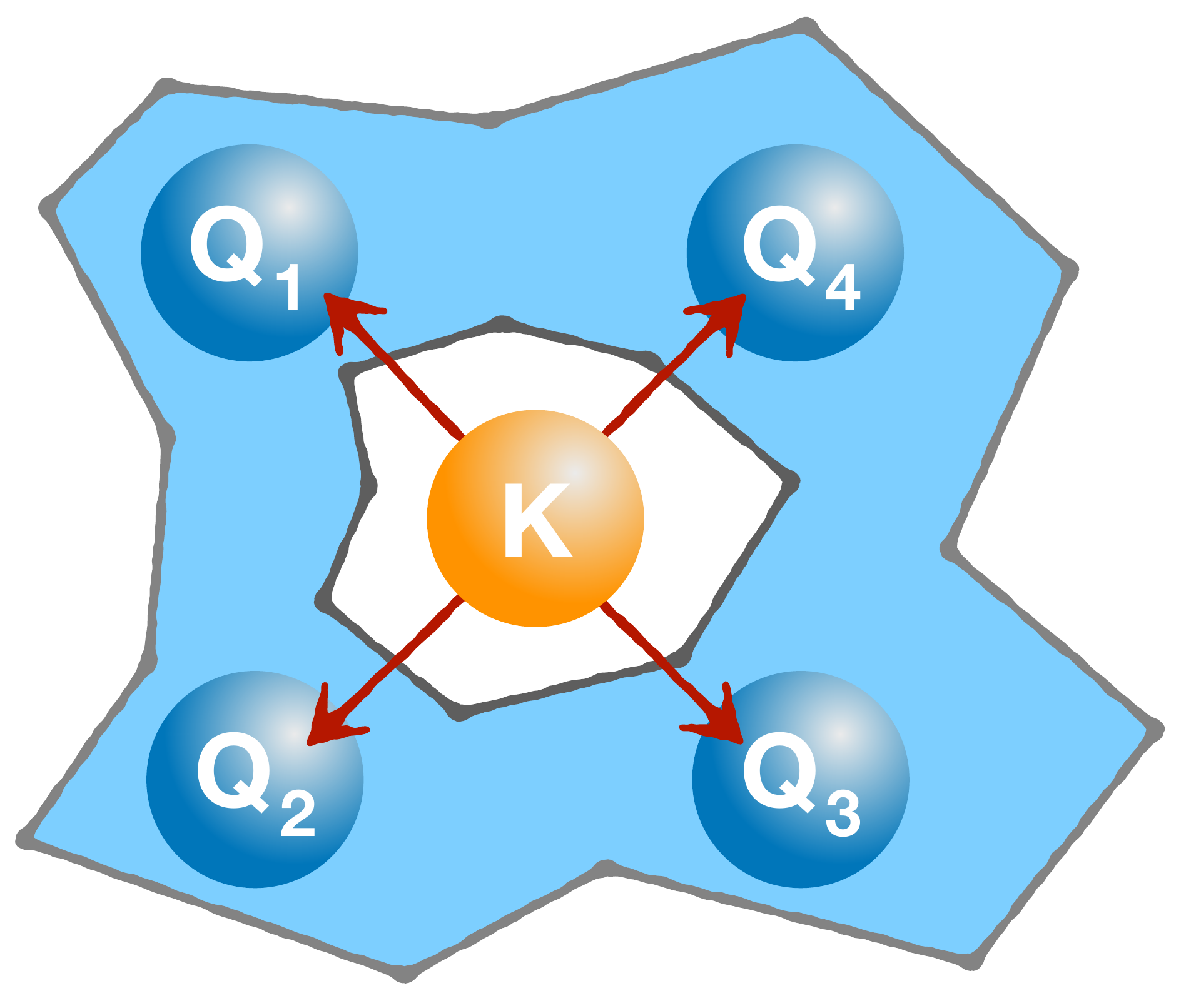}\\
    {\bf (b)}
    \includegraphics[width=0.75\columnwidth]{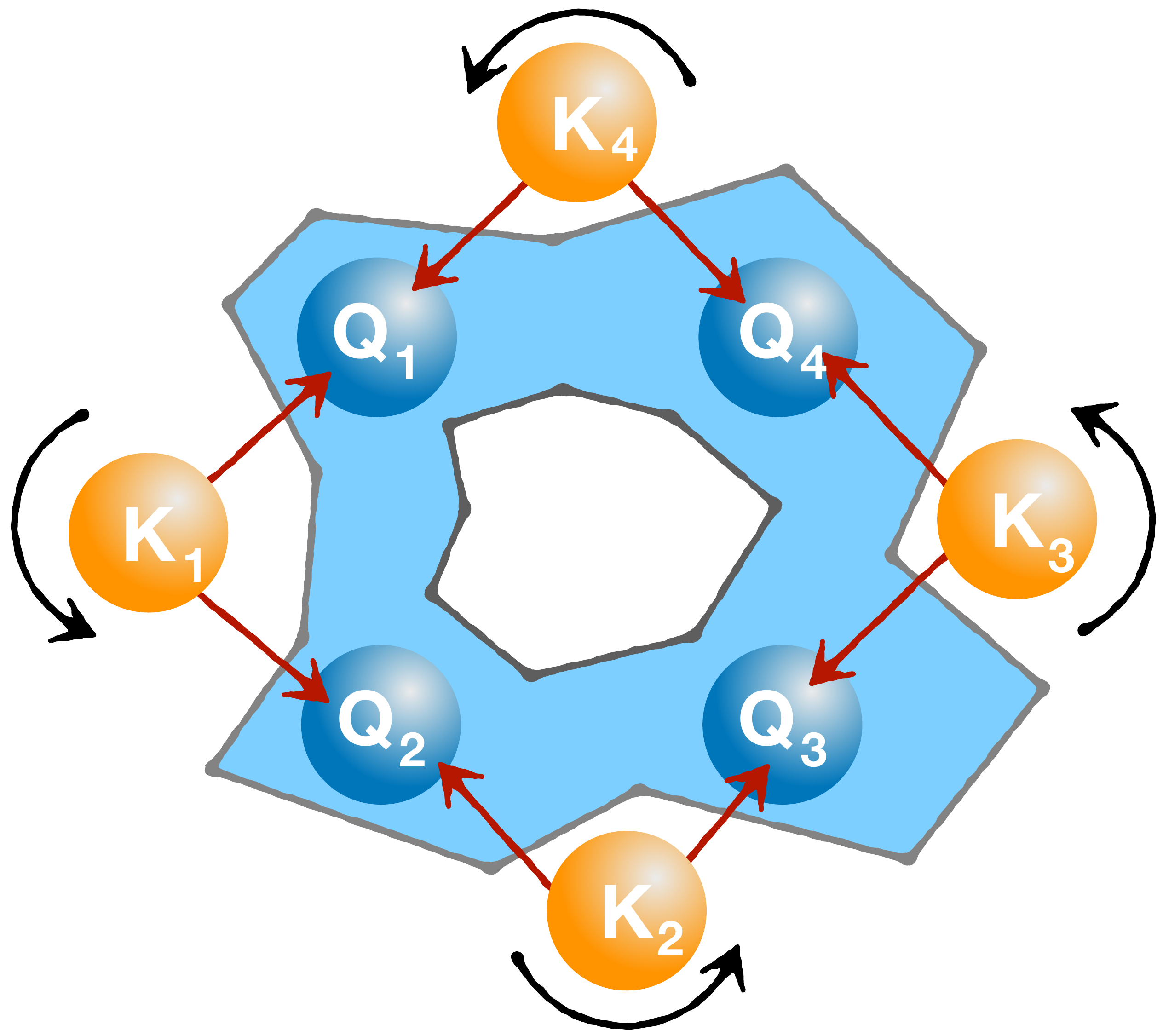}
    \caption{\textbf{Ring topology protocols.} {\bf (a)} Single-carrier protocol: the carrier qubit $K$ interacts once with each node; {\bf (b)} qudit carrier case: the carrier travels in one direction, interacting twice with each node, once for encoding and the second time for decoding, since the encoding and decoding steps involve different subspaces of the carrier.}
    \label{fig:ring}
\end{figure}

We investigate a four-node case where the qubits $Q_{1,..,4}$ are entangled as a result of the application of the protocol illustrated before. As we request explicitly that $Q_1$ and $Q_4$ are entangled, we would thus realize a {\it ring-like} structure [cf. Fig.~\ref{fig:ring} {\bf (a)}]. 
The initial state of the nodes, then, must include non-classical correlations between every possible pair $\{{\cal C}_k\}_{k=1}^4$, where ${\cal C}_k=\{Q_k,Q_{k+1}\}$ and we set $Q_5=Q_1$, so that
\begin{equation}
\begin{split}
  \alpha_4=\frac{1}{4}\big(\rho^0_{Q_1,Q_2}\otimes\alpha^0_{Q_3}\otimes\alpha^0_{Q_4}+\rho^0_{Q_2,Q_3}\otimes\alpha^0_{Q_1}\otimes\alpha^0_{Q_4}+\\+\rho^0_{Q_3,Q_4}\otimes\alpha^0_{Q_1}\otimes\alpha^0_{Q_2}+\rho^0_{Q_4,Q_1}\otimes\alpha^0_{Q_2}\otimes\alpha^0_{Q_3}\big)  
\end{split}
\end{equation}
where $\rho^0_{{\cal C}_k}$ and $\alpha^0_{Q_i}$ are the same as in Eq. \eqref{eq:initialAlphaN}.
The protocol consists of only four steps, taking the initial state $\alpha_4$ to the final one as 
\begin{equation}
\eta_4=\left(\Pi^4_{j=1}\mathcal{P}_{Q_jK}\right)
(\alpha_4\textcolor{black}{\otimes \alpha_K})\left(\Pi^4_{j=1}\mathcal{P}^\dag_{Q_jK}\right).
\end{equation}

In particular, the amount of entanglement in each of the one-vs-four bipartitions of the form $Q_j|{\cal G}_K$ with ${\cal G}_K=\{Q_1,Q_2,Q_3,Q_4,K\}\setminus{Q_j}$ that can be identified in state $\eta_4$ is the same: the corresponding partially transposed density matrices $\eta^{\text{PT}_{Q_j}}$ all have a single negative eigenvalue equal to -0.0175206, so that the entanglement $\mathcal{E}_{Q_j|{\cal G}_K}$ does not depend on $j=1,..,4$. 
 On the other hand, the entanglement 
 $\mathcal{E}_{K|Q_{1,..,4}}$ between the carrier $K$ and the network is identically zero, thus achieving a successful distribution of entanglement without involving the carrier.
\textcolor{black}{It is worth noting that after projecting the carrier system onto state $\ket{A}$ and tracing the carrier system away we obtain a reduced matrix of the network only which exhibit the same entanglement values.}

In order to demonstrate that the system actually features multipartite entanglement, we check the eigenvalues of the partial transpose with respect to any possible bipartition of the system. As reported in Tab. \ref{tab:ring_eigenvalues}, we fulfill this requirement. 

\subsubsection{Multi-qubit carrier}
We now move to the study of a multi-carrier configuration, and how this might affect the effectiveness of the protocol. As in the ring pattern we have to {weave} four entanglement links. We thus 
consider a compound carrier system of 4 qubits $K_j~(j=1,..,4)$. The protocol differs from the single-carrier one in the exploitation of different carrier subspaces for the encoding and decoding operations affecting different node pairings. This implies that each operation will only act on a certain link, depending on the nodes that are involved.
Therefore, the protocol needs twice the number of steps required in the single-qubit carrier scheme. Such steps are explicitly illustrated in Table~\ref{tab1}.
\begin{table}[b]
     \begin{tabular}{c|c|c}
    \text{State}&\text{Description}&\text{Encoder/}\\
    \text{label}&\text{of evolution}&\text{Decoder}\\
    \hline
    \hline
        $\beta_T$ & $\mathcal{P}_{Q_1 K_1}\alpha_T\mathcal{P}_{Q_1 K_1}^+$ & $K_1$ \\
        $\gamma_T$ & $\mathcal{P}_{Q_2 K_1}\beta_T\mathcal{P}_{Q_2 K_1}^+$& $K_1$ \\
       $ \delta_T$&$\mathcal{P}_{Q_2 K_2}\gamma_T\mathcal{P}_{Q_2 K_2}^+$&$K_2$\\
        $\eta_T$&$\mathcal{P}_{Q_3 K_2}\delta_T\mathcal{P}_{Q_3 K_2}^+$&$K_2$\\
        $\zeta_T$&$\mathcal{P}_{Q_3 K_3}\eta_T\mathcal{P}_{Q_3 K_3}^+$&$K_3$\\
        $\kappa_T$&$\mathcal{P}_{Q_4 K_3}\zeta_T\mathcal{P}_{Q_4 K_3}^+$&$K_3$\\
        $\chi_T$&$\mathcal{P}_{Q_4 K_4}\kappa_T\mathcal{P}_{Q_4 K_4}^+ $&$K_4$\\
        $\omega_T$&$\mathcal{P}_{Q_1 K_4}\chi_T\mathcal{P}_{Q_1 K_4}^+$&$K_4$
        \end{tabular}
            \caption{Description of the steps required in a multi-qubit carrier protocol. We provide the label of the state achieved at each step of the scheme, the corresponding encoding (decoding) operation and the associated encoder (decoder).}
              \label{tab1}
        \end{table}


We report a sketch of the procedure in Fig. \ref{fig:ring} {\bf (b)}.
We compute again the eigenvalues for any possible bipartition of the system, reporting them in Tab. \ref{tab:ring_eigenvalues}.
The results of our analysis show that also a qudit carrier approach produces multipartite entanglement. 
A main drawback comes from the fact that, although the carrier is in a separable state, tracing it away presents some complications.
Since each of the different entanglement links is mediated by a different subspace, the projection of the qubit carrier on the state $\ket{A}$ will result in the preservation of that link in the reduced network state. Unfortunately, that can't be done simultaneously for all the node pairs: by projecting every qubit carrier on their respective $\ket{A}$ state, we get a separable reduced state of the network. Therefore, the final state for the system remains multipartite entangled as far as the carrier state is not further manipulated. The carrier can be only traced away in case we wish to observe a specific entanglement link between two nodes.

\subsubsection{General remarks on ring topology protocols}

In order to compare the effectiveness of the two protocols we compute the average negativity of the final state for both cases.
Given a certain partition $p$ of a composite state $\rho$, we can define negativity as in \cite{vidal2002computable}
\begin{equation}
    \mathcal{N}_p(\rho)=\frac{||\rho^{T_p}||-1}{2}
\end{equation}
which is equal to the sum of all negative eigenvalues of the transposition of $\rho$ with respect to the partition $p$.
 We consider the geometrical average of the negativity values for all the possible partitions of the system, denominating this value $\mathcal{N}$. We obtain a $\mathcal{N}=0.0184179$ ($\mathcal{N}=0.0261631$) for the single qubit carrier (multi-qubit carrier) protocol.
 These results show some advantage coming from the employment of a high-dimensional carrier in terms of entanglement production, though implying, in the perspective of an experimental realization, far heavier efforts and drawbacks.
 
\begin{figure}[b]
    \centering
    \includegraphics[scale=0.55]{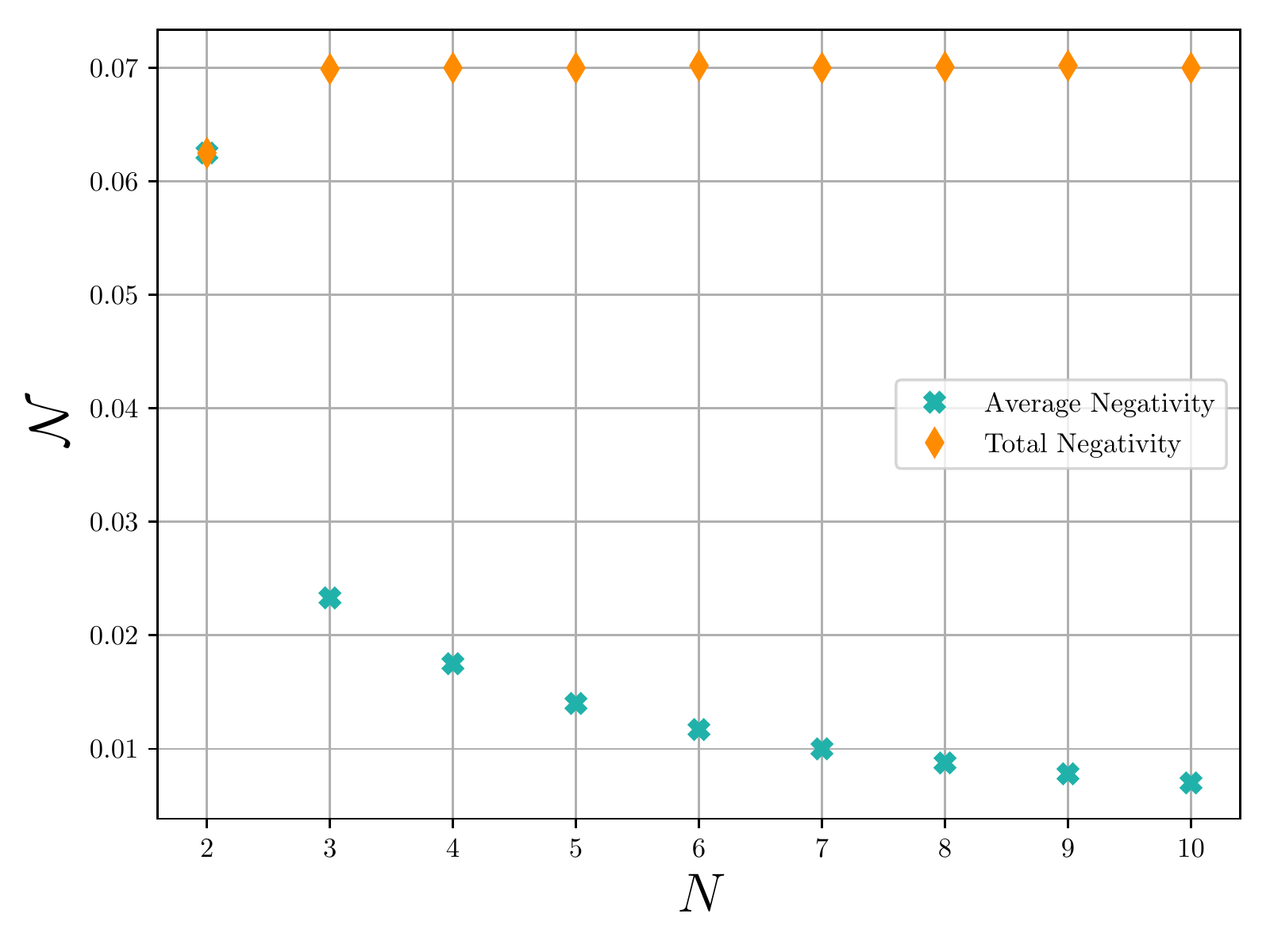}
    \caption{\textbf{Negativity vs number of network nodes.} 
    The average negativity $\mathcal{N}$ is reported in function of the number of nodes $N$ involved in the entanglement distribution protocol together with the corresponding total negativity. The investigation is carried for the single qubit carrier case, in the ring pattern scenario.}
    \label{fig:neg_trend}
\end{figure}

 It is worth noting that the amount of entanglement produced on average for a single link is lower than in the binary case of Ref.~ \cite{fedrizzi2013experimental}. This is understandable considering the fact that the 4-nodes initial mixed state contains many more terms which generate ``noise" contributions in the final state, with respect to the 2 nodes case. Indeed, we expect the average produced negativity to decrease as the number of nodes increases, together with the number of terms to be included in the initial state. It may even be possible that, after a certain size of the network, entanglement between the nodes is no longer detectable.
 We investigate the ring pattern case up to $N=10$ nodes, in order to understand the trend of negativity in function of the size of the network, reporting the results in Fig. \ref{fig:neg_trend}.
 \textcolor{black}{Our simulations confirm the expected decrease of the average negativity, but also highlights that the total negativity remains constant as we increase the number of entangled nodes; therefore, the addition of more nodes does not seem to jeopardize the protocol efficiency in converting discord into entanglement. }

\begin{figure}[h]
    \includegraphics[width=0.75\columnwidth]{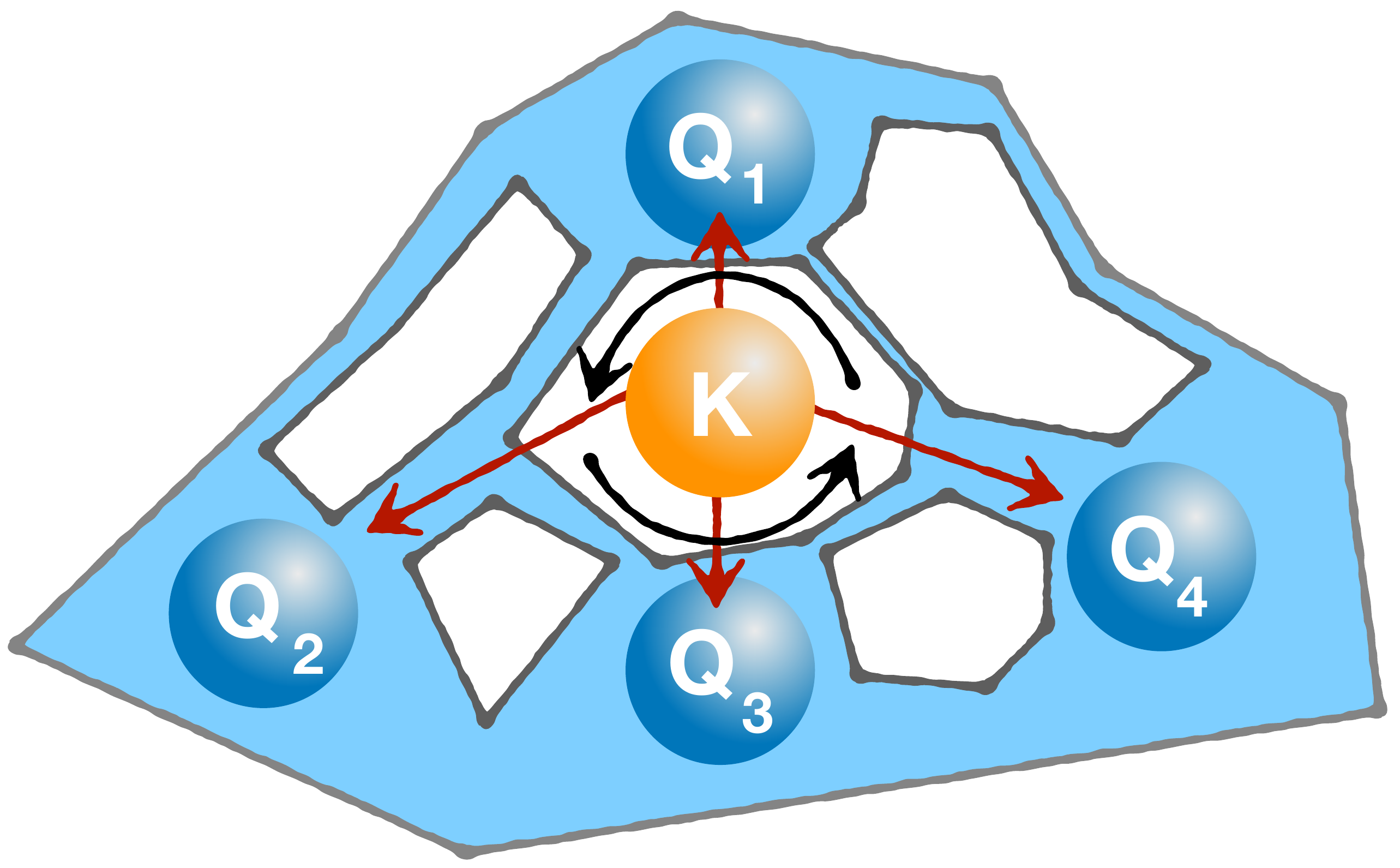}
    \caption{\textbf{Star topology protocol.} Single qubit carrier case: the qubit travels in one direction, interacting once with each node. We refer to the main body of the manuscript for a description of the qudit carrier case. 
    }
    \label{fig:star}
\end{figure}

\subsection{Four nodes example: Star Topology}

In order to provide a more thorough analysis of the potentialities of our approach, we tailor the protocol to generate a star-like entanglement pattern. We design the initial state and the protocol steps with the aim of producing a final state in which one central node is entangled with all the others. In this case, the final state results in entanglement with respect to any possible bipartition of the system. We briefly report on this analysis, because of the many analogies with the ring pattern case.
We consider four nodes ${\cal Q}=\{Q_1,Q_2,Q_3,Q_4\}$, with $Q_1$ as the central node. Hence, our initial state has to be the mixture of three terms, each featuring non-classical correlation between qubit $Q_1$ and the others: 
\begin{equation}
\alpha_4=\frac{1}{3}\sum^4_{j=2}\rho^0_{{Q_1},{Q_j}}\otimes\left(\bigotimes_{\overline{j}\in\overline{\cal Q}}{\rho_Q}_{\overline{j}}\right)
\end{equation}
with $\overline{\cal Q}={\cal Q}\backslash\{Q_1,Q_{j}\}$. 
In this case, the single qubit carrier protocol proceeds identically to the ring case: the carrier interacts at first with the central node and then once with each other qubit, as depicted in Fig. \ref{fig:star}, 
since the first operation acts as encoding for every entanglement link. The only difference consists of the initial state preparation of the network, a remarkable feature in terms of flexibility of our strategy.

The qudit carrier case is more complex: each encoding and decoding operation have to be addressed separately, having the nodes interact with different sub-qubits of the carrier. Therefore, the carrier has to travel back and forth from the central node to the periferic qubits, until each link has been woven. Considering a eight-dimensional qudit carrier and three qubit subsystems $K_1$, $K_2$ and $K_3$, we can explicitly write down the protocol:
\begin{equation}
\begin{split}
    \beta_T=\mathcal{P}_{Q_1 K_1}\alpha_T\mathcal{P}_{Q_1 K_1}^+ \text{encoding mediated by $K_1$}\\
    \rightarrow \gamma_T=\mathcal{P}_{Q_2 K_1}\beta_T\mathcal{P}_{Q_2 K_1}^+ \text{decoding mediated by $K_1$}\\
    \rightarrow \delta_T=\mathcal{P}_{Q_1 K_2}\gamma_T\mathcal{P}_{Q_1 K_2}^+ \text{encoding mediated by $K_2$}\\
    \rightarrow \eta_T=\mathcal{P}_{Q_3 K_2}\delta_T\mathcal{P}_{Q_3 K_2}^+ \text{decoding mediated by $K_2$}\\
    \rightarrow \zeta_T=\mathcal{P}_{Q_1 K_3}\eta_T\mathcal{P}_{Q_1 K_3}^+  \text{encoding mediated by $K_3$}\\
    \textcolor{black}{\rightarrow \kappa_T=\mathcal{P}_{Q_4 K_3}\zeta_T\mathcal{P}_{Q_4 K_3}^+  \text{decoding mediated by $K_3$}}.
\end{split}
\end{equation}
We report in Tab. \ref{tab:star_eigenvalues} the negative eigenvalues relative to every bipartition of the system for both methods.
The average negativity computed from these results reads $\mathcal{N}=0.019268$ for the qubit carrier protocol and $\mathcal{N}=0.0262659$ for the qudit carrier one. In this case, the gap in entanglement production due to the exploitation of a high-dimensional carrier is \textcolor{black}{slightly lower} 
with respect to the ring pattern case, while the other issues remain.
In general, the comparison between the usage of a qubit or a qudit carrier may provide different answers according to the application case and, more importantly, the actual experimental situation we are dealing with.\\

\section{Experimental proposals} \label{sec:experiment}

We propose some feasible experimental ways of demonstrating the effectiveness of our protocol in an optical framework.

\subsubsection{Single qubit carrier}

\begin{figure}[t]
    \centering
    \includegraphics[width=\columnwidth]{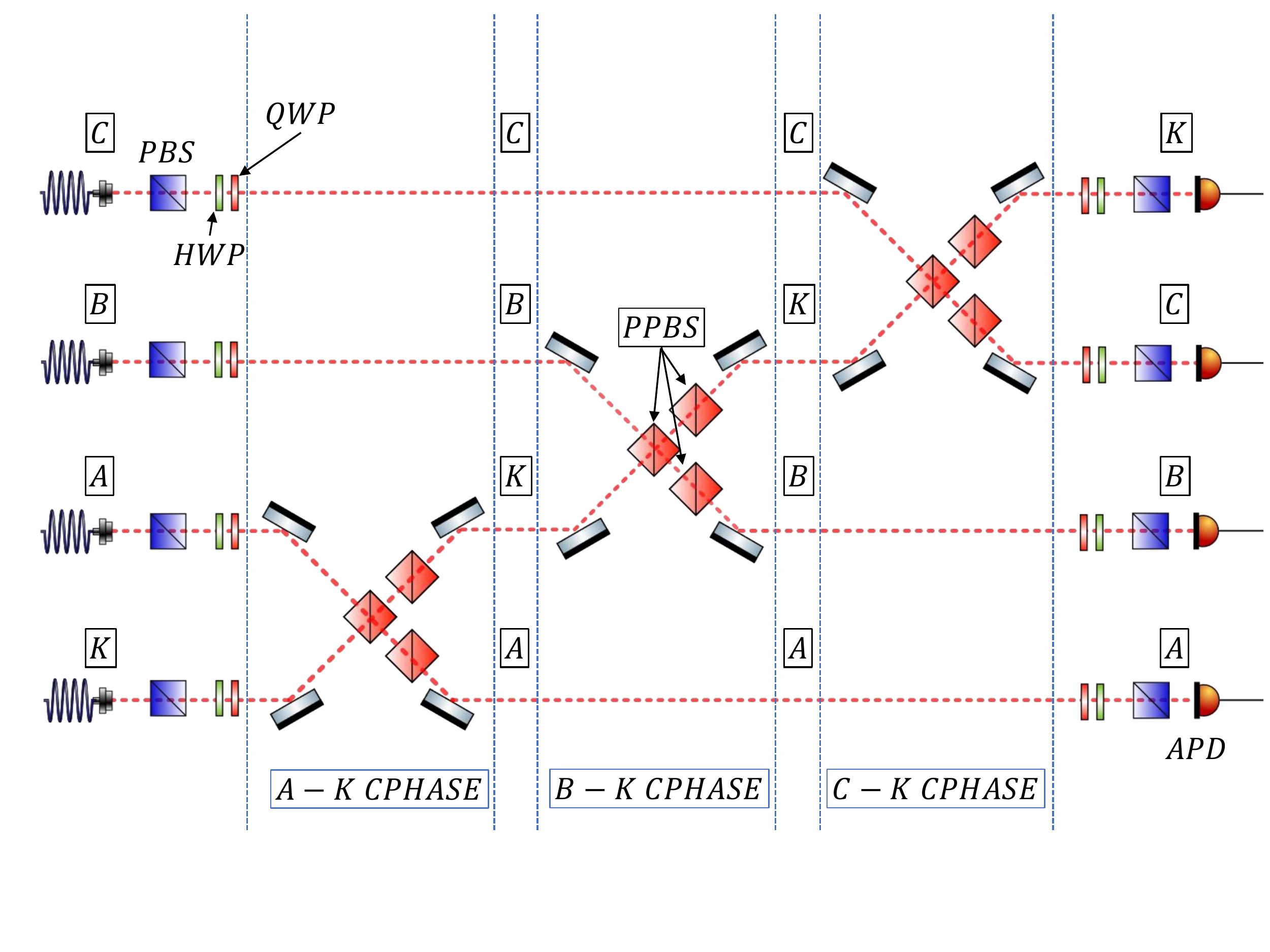}
    \caption{\textbf{Scheme of principle for the Distribution Protocol for a 3 nodes ring pattern.} {Four indistinguishable photons must be harnessed, with a likely necessary further one for heralding. Three of them act as nodes and interact in turn with the fourth photon, acting as the carrier qubit. Initial state preparation in the degree of freedom of polarization can be addressed through the sequence of optical elements: Polarizing Beamsplitter, Half-wave Plate and Quarter-wave plate. The interaction between qubits can be implemented via the optical CPHASE gate described in \cite{langford2005entangling,fedrizzi2013experimental}. The sequence of physical interactions between the photons shall yield as a result a ring-like multipartite entanglement structure, or even a star-like structure, according to the initial state being implemented.}}
    \label{fig:setup_1}
\end{figure}

The direct experimental implementation of the single qubit carrier may well be a direct generalization of the apparatus of \cite{fedrizzi2013experimental}: $N$ single photons are employed, one as a carrier qubit, while the others act as the network nodes. The state of the network is encoded in the polarization degree of freedom of photons. All photons have to be indistinguishable with the carrier (hence reciprocally indistinguishable) in order to implement the optical quantum CZ gate as described in \cite{langford2005entangling,fedrizzi2013experimental}, which acts as the encoding/decoding operation.
That may be very difficult to obtain for a high number of photons: they have to be synchronized and identical in any degree of freedom. Indeed, it is possible to build sources with a such a control on the photon generation, which allow many photons interaction \cite{jones2020interference}.
In Fig. \ref{fig:setup_1}, we report a sketch of the possible experimental implementation of the protocol for the ring pattern for three nodes. After each encoding/decoding operation the photon acting as carrier is sent to the next node and interacts with the corresponding photon, until it has interacted with all the network nodes and it can be projected and measured, leaving, in principle, an entangled state of the network.
The most complex part of the protocol resides in the state preparation, but, if we wish to provide a mere experimental demonstration of the protocol effectiveness, the mixing probabilities of the various terms in the initial state may be simulated by different sampling times, as already done in \cite{fedrizzi2013experimental}. That couldn't be the case in actual application scenarios.

\subsubsection{Single qubit carrier, relay scheme}

\begin{figure}[h]
    \centering
    \includegraphics[scale=0.5]{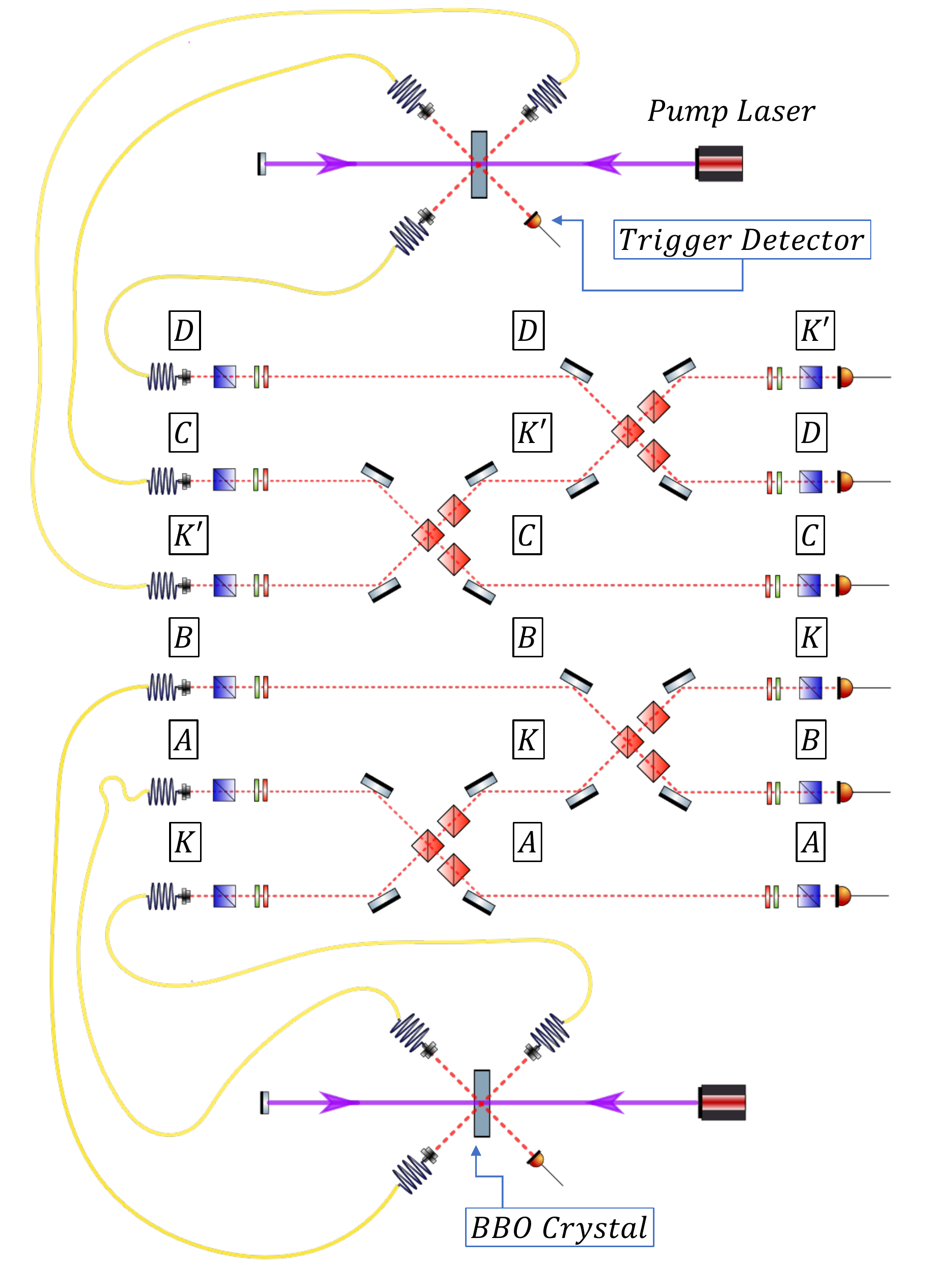}
    \caption{\textbf{Scheme of principle for the Relay Protocol for a 4 nodes ring pattern.} {Two BBO crystals pumped by a pulsed laser generate four indistinguishable photons altogether. Three of them are sent into the setup, while the fourth is used to witness as a herald for generation. Two sources allow to generate two triplets of indistinguishable photons, each of which can be used to address two nodes and one carrier. The photons are prepared in the suitable initial state after traveling through an optical fiber, then three of them are used to perform the encoding and decoding of entanglement between $A$ and $B$ via $K$. After the measurement of $K$, a second triplet can be used to weave the entanglement  between $C$ and $D$ via $K'$ and, concurrently, the links between $B$ and $C$ and $D$ and $A$.}}
    \label{fig:setup_2}
\end{figure}

The request of producing $N$ indistinguishable photons, even for low $N>3$, may actually be difficult to comply with. 
There is another suitable way to demonstrate the protocol in an experimental implementation: in Appendix B we report on a variation of the protocol, relying on a ``relay scheme", where the single qubit carrier is measured and replaced after a certain number of protocol steps.
This scheme is less efficient than the standard one, but possesses some interesting features as we face an experimental realization.
Indeed, the relay strategy implies that the carrier needs only to be indistinguishable with the qubit nodes it interacts with. In particular, we consider the case in which a new qubit carrier is employed after the previous one has interacted with two nodes. Therefore, again in an optical framework, we only need to generate three indistinguishable photons, plus a triggering one, which corresponds to what has been already realized in \cite{fedrizzi2013experimental} for the binary protocol. The $N$ nodes protocol may be realized by exploiting $N/2$ sources in an actual scenario.\\
In case of a proof-of-principle framework, it would be even possible to use the same photon source, since the different parts of the systems remain completely isolated throughout the protocol. It would only be necessary to suitably set the photons' state as the qubit pair which is simulated to be under observation.  
A sketch of the possible experimental implementation of the scheme for the case of the ring pattern in four nodes is reported in Fig. \ref{fig:setup_2}.

\section{Concluding Remarks} \label{sec:conclusions}
We have presented a scheme for the achievement of EDSS in a  multipartite network. In contrast with the proposal in Ref.~\cite{karimipour2015systematics}, our strategy is a generalization of the one proposed in \cite{fedrizzi2013experimental} and so does not need supplemental manipulation of the carrier system, aimed at disentangling it.
It is also characterized by a remarkable flexibility in terms of feasible distribution patterns and possible variations of the standard scheme.
Indeed, our strategy may be extended to a continuous-variable framework, as in the binary case, in order to make experimental implementations easier to realize.
The results of our work provide a very general alternative approach to direct protocols in the problem of entanglement distribution, although, as highlighted in the text, it is a probabilistic approach. Nevertheless, the advantages of using a separable state carrier in some environmental conditions \cite{fedrizzi2013experimental}, may be finally extended to the $N$ qubits scenario, where noise can play a very relevant part. Therefore, these results may pave the way to the development of the general and useful application of EDSS protocols in actual multiparty Quantum Communication and Information tasks.

For instance, an application of this work worth highlighting is quantum conference key agreement (QCKA)~\cite{murta20quantum}, that is, the ability to use properties of quantum states to securely share secret keys between $N>2$ parties. Quantum key distribution is becoming increasingly important as we approach the realisation of quantum computers which would render existing security protocols useless. In particular, QCKA is growing in relevance as quantum networks of many nodes are being developed for the purpose of secure communication (for instance, see Refs.~\cite{liao18satellite,dynes19cambridge,aguado19engineering}).

Remarkably, the resources for QCKA are precisely states with the entanglement structure achieved through the protocol illustrated here~\cite{carrara21genuine}. Once the entangled state has been shared among the desired $N$ nodes according to the protocol in Sec.~\ref{sec:protocol}, the N-BB84 protocol~\cite{grasselli18finitekey,murta20quantum} for QCKA could then be used to establish a secret key. Therefore, not only can we distribute keys securely between $N$ parties without needing to send, for instance, fragile GHZ states to several nodes, but we show that it is in fact possible without sending any entanglement at all. 

\acknowledgments
We wish to thank professor Paolo Mataloni for facilitating and supporting the visit of A.L. to the Centre for
Theoretical Atomic, Molecular and Optical Physics during which this work has started.
We acknowledge support from the European Union's Horizon 2020 FET-Open project TEQ (766900), the Leverhulme Trust Research Project Grant UltraQuTe (grant RGP-2018-266), the Royal Society Wolfson Fellowship (RSWF/R3/183013), the UK EPSRC (grant EP/T028424/1) and the Department for the Economy Northern Ireland under the US-Ireland R\&D Partnership Programme. 

\section*{Appendix A: simultaneously entangled pairs}

At the end of Section III, we described the form of the final state of the network as a mixture of terms featuring entanglement between a single node pair each. Clearly, the initial state can be tailored to obtain different mixtures.
We show this via a ring pattern example, where we consider a network of $N$ nodes $\{Q_i\}_{i=1}^N$, where we consider $N$ to be even for sake of simplicity. We want to obtain a final state of the network featuring multipartite entanglement and the possibility of having simultaneously entangled node pairs in the system.
We set the initial state as the mixture of two terms
\begin{equation}
\alpha_N=\frac{1}{2}\left(\bigotimes_{k=1}^{N/2}\rho_{Q_{2k-1},Q_{2k}}+\bigotimes_{j=0}^{N/2-1}\rho_{Q_{2j},Q_{2j+1}}\right)
\end{equation}
where we consider $Q_0=Q_N$. In this case, simply applying the single qubit carrier protocol, we would obtain a final state of the network fulfilling our initial requirements, but the carrier would end up being entangled during the process.
Therefore, we mix these two terms with the initial state for a $N$ nodes ring-like entanglement distribution pattern
\begin{equation}
\begin{split}
\alpha_N&=\frac{1}{N+2}\left[\bigotimes_{k=1}^{N/2}\rho_{Q_{2k-1},Q_{2k}}+\bigotimes_{j=0}^{N/2-1}\rho_{Q_{2j},Q_{2j+1}}+\right.\\
&+\left.\sum_{k=1}^N \left(\bigotimes_{i=1}^k\ket{0}\bra{0}_{Q_i}\right)\otimes\rho^0_{Q_k,Q_{k+1}}\otimes\left(\bigotimes_{j=k+2}^N\ket{0}\bra{0}_{Q_j}\right)\right]
\end{split}
\end{equation}
where we consider $Q_{N+1}=Q_1$.
In this way, we are inserting ``noise" in the state of the system, which helps keep the carrier separable, while diminishing the probability of finding the network in a final state featuring simultaneously entangled node pairs.
After the application of the single-qubit carrier protocol, the final state will be
\begin{equation}
\begin{aligned}
    &\rho^f_N=p\Omega_N
    +q\left(\sum_{k=1}^N \ket{\phi^+}\bra{\phi^+}_{Q_i,Q_{i+1}}\right)\otimes\left(\bigotimes_{j=1, j\neq i,i+1}^{N-1}\alpha^0_{Q_j}\right)\\
    &+r\left(\bigotimes_{k=1}^{N/2}\ket{\phi^+}\bra{\phi^+}_{Q_{2k-1},Q_{2k}}+\bigotimes_{j=0}^{N/2-1}\ket{\phi^+}\bra{\phi^+}_{Q_{2j},Q_{2j+1}}\right)
    \end{aligned}
\end{equation}
where $p,q,r\in\mathcal{R}$ with $p+q+r=1$, and $\Omega_N$ is again a completely diagonal contribution to the state of the network, hence classical.
Therefore, we have a certain probability of actually finding the system in the state we desire.
Alternatively, if we relax the request of multipartite entanglement, we can use the initial state $\alpha=\bigotimes_{k=1}^{N/2}\rho_{Q_{2k-1},Q_{2k}}$, relatively increasing the probability of finding the network in a product state of entangled node pairs, although the system shall remain separable with respect to some bipartitions.

\section*{Appendix B: relay scheme for single qubit carrier}
It is possible to define a single qubit carrier scheme in which the carrier is halfway replaced with another qubit.
This alternative protocol may result very useful for possible future experimental implementations, as highlighted in Section~\ref{sec:experiment}.
We explicitly show the details of this relay scheme in a four nodes ring pattern example.
We have four nodes $A, B, C, D$ and a carrier qubit $K$.
The initial state of the network and carrier is the same as in the standard ring pattern case $\alpha_T$, as well as the encoding and decoding operations. The only difference consists of the fact that after two interactions, we project the qubit system on the $\ket{A}\bra{A}$ state, we trace it away and we insert a new qubit carrier in the initial state $\alpha_K$
\begin{equation}
\begin{aligned}
    \beta_T&=\mathcal{P}_{AK}\alpha_T\mathcal{P}_{AK}^+\\
    \rightarrow \gamma_T&=\mathcal{P}_{BK}\beta_T\mathcal{P}_{BK}^+\\
    \rightarrow \gamma_T^{'}&=\ket{A}_K\bra{A}\gamma_T\ket{A}_K\bra{A}\\
    \rightarrow \gamma_N&=Tr_K(\gamma_T^{'})\\
    \rightarrow\gamma_T^{''}&=\gamma_N\otimes\alpha'_K\\
    \rightarrow \delta_T&=\mathcal{P}_{CK^{'}}\gamma^{''}_T\mathcal{P}_{CK^{'}}^+\\
    \rightarrow \eta_T&=\mathcal{P}_{DK^{'}}\delta_T\mathcal{P}_{DK^{'}}^+
\end{aligned}
\end{equation}

At the end of the protocols, we get the following negative eigenvalues from the partial transpositions of $\eta_T$
\begin{equation}
    \begin{cases}
    \mathcal{E}_{A-BCDK^{'}}=\{-0.00986842, -0.00328947\}\\
    \mathcal{E}_{B-ACDK^{'}}=-0.00986842\\
    \mathcal{E}_{C-ABDK^{'}}=\{-0.00986842, -0.00328947\}\\
    \mathcal{E}_{D-ABCK^{'}}=-0.00986842\\
    \mathcal{E}_{K^{'}-ABCD}=0
    \end{cases}
\end{equation}
and by analyzing all the partition the system exhibits multipartite entanglement. It is quite evident that the average negativity produced is lower than in the standard case, although the requested multipartite entanglement and carrier separability are achieved.
Therefore, the relay scheme provides with a weaker yet effective protocol for EDSS, which may prove to be useful in practical applications.\\

\section*{Appendix C: Tables of eigenvalues}
\label{appendixC}

\newcommand{\resetcount}{\renewcommand{\theequation}{C-\arabic{equation}}
\setcounter{equation}{0}}

\begin{table*}[t!]
\centering
\begin{tabular}{c|c|c}
\hline
    Bipartition & Single-carrier & Multiple-carrier\\
    \hline\hline
    $Q_1|Q_2Q_3Q_4K$&-0.0175206 & $\{-0.011786, -0.00392868,-0.00392868, -0.001309565\}$ \\
    \hline
    $Q_2|Q_1Q_3Q_4K$&-0.0175206 & $\{-0.011786, -0.00392868,-0.00392868, -0.001309565\}$ \\
    \hline
    $Q_3|Q_1Q_2Q_4K$&-0.0175206 & $\{-0.011786, -0.00392868,-0.00392868, -0.001309565\}$ \\
    \hline
    $Q_4|Q_1Q_2Q_3K$&-0.0175206 & $\{-0.011786, -0.00392868,-0.00392868, -0.001309565\}$ \\
    \hline
    $Q_1Q_2|Q_3Q_4K$&$\{-0.0078125, -0.0078125\}$ & $\{-0.00769043, -0.00769043, -0.00286949, -0.00256348,$\\ 
    & & $-0.00256348, -0.00256348, -0.00256348, -0.000956497,$ \\
    & & $-0.000956497, -0.000854492, -0.000854492, -0.000318832\}$ \\
    \hline
    $Q_1Q_3|Q_2Q_4K$&$\textcolor{black}{-0.03125}$& $\{-0.0117871, -0.0117871, -0.00395737,-0.00395737,$ \\
    & & $-0.00395737,-0.00395737, -0.00195313\}$\\
    \hline
    $Q_1Q_4|Q_2Q_3K$&$\textcolor{black}{\{-0.0078125, -0.0078125\}}$ & $\{-0.00769043, -0.00769043, -0.00286949, -0.00256348, $ \\
    & & $-0.00256348, -0.00256348, -0.00256348, -0.000956497,$\\
    & & $0.000956497, -0.000854492, -0.000854492, -0.000318832\}$\\
    \hline
\end{tabular}
\caption{{Negative eigenvalues of every possible partition, ring pattern, qubit and qudit carrier protocols}}
\label{tab:ring_eigenvalues}
\end{table*}


\begin{table*}[t!]
\centering
\begin{tabular}{c|c|c}
\hline
    Bipartition & Single-carrier & Multiple-carrier\\
    \hline\hline
    $Q_1|Q_2Q_3Q_4K$&$-0.0342865$& $\{-0.0291511, -0.00642872,-0.00642872, -0.00642872\}$ \\
    \hline
    $Q_2|Q_1Q_3Q_4K$&$-0.0121071$& $\{-0.00681022, -0.00227007,-0.00227007, -0.000756691\}$ \\
    \hline
    $Q_3|Q_1Q_2Q_4K$&$-0.0121071$ & $\{-0.00681022, -0.00227007,-0.00227007, -0.000756691\}$ \\
    \hline
    $Q_4|Q_1Q_2Q_3K$&$-0.0121071$ & $\{-0.00681022, -0.00227007,-0.00227007, -0.000756691\}$ \\
    \hline
    $Q_1Q_2|Q_3Q_4K$&$-0.0245719$ & $\{-0.0235657, -0.00681022,-0.00460722,$\\ 
    & & $-0.00460722,-0.00460722, -0.00227007\}$ \\
    \hline
    $Q_1Q_3|Q_2Q_4K$&$\textcolor{black}{ -0.0245719}$& $\{-0.0235657, -0.00681022,-0.00460722,$ \\
    & & $-0.00460722, -0.00227007\}$\\
    \hline
    $Q_1Q_4|Q_2Q_3K$&$\textcolor{black}{ -0.0245719}$ & $\{-0.0235657, -0.00681022, -0.00460722,  $ \\
    & & $-0.00460722, -0.00460722, -0.00227007\}$\\
    & & \\
    \hline
\end{tabular}
\caption{{Negative eigenvalues of every possible partition, star pattern, qubit and qudit carrier protocols}}
\label{tab:star_eigenvalues}
\end{table*}


\bibliography{edss}

\end{document}